\documentclass[aps,twocolumn,pra,showpacs,amsmath,amssymb,showkeys,10pt]{revtex4-2}
\usepackage{graphicx}
\usepackage{epstopdf}
\pdfoutput=1
\usepackage{braket}
\usepackage{hyperref}
\usepackage{longtable}

\begin{document}

	\title{Entanglement and quantum discord in the cavity QED models}
	
	\author{Miao Hui-hui}
	\email[Email address: ]{hhmiao@cs.msu.ru}
	\affiliation{Faculty of Computational Mathematics and Cybernetics, Lomonosov Moscow State University, Vorobyovy Gory 1, Moscow, 119991, Russia}

	\author{Li Wan-shun}
	\email[Email address: ]{wanshun.li@cs.msu.ru}
	\affiliation{Faculty of Computational Mathematics and Cybernetics, Lomonosov Moscow State University, Vorobyovy Gory 1, Moscow, 119991, Russia}

	\date{\today}

	\begin{abstract}
	We investigate the quantum correlation between light and matter in bipartite quantum systems, drawing on the Jaynes--Cummings model and the Tavis--Cummings model, which are well-established in cavity quantum electrodynamics. Through the resolution of the quantum master equation, we can derive the dissipative dynamics in open systems. To assess the extent of quantum correlation, several measures are introduced: von Neumann entropy, concurrence and quantum discord. The effects of initial entanglement and dissipation intensity on quantum discord are carefully examined. Furthermore, we examined the dynamics of quantum discord within the $\rm{OH}^+$ model.
	\end{abstract}


	\maketitle

	\section{Introduction}
	\label{sec:Introduction}
	
	The concept of quantum entanglement \cite{Horodecki2009} was inspired by the Einstein--Podolsky--Rosen (EPR) paradox, as revealed by Einstein et al \cite{Einstein1935}. Quantum entanglement has consistently been a significant area of investigation in fundamental aspects of quantum information theory. In contrast to the conventional approach, quantum information processing (QIP), linked with entanglement, provides an enhanced spectrum of information manipulation capabilities. In the context of QIP, including quantum computation \cite{Nielsen2010}, quantum cryptography \cite{Ekert1991}, and additional studies \cite{Bennett1992, Bennett1993, Miao2024}, quantum entanglement serves as a distinctive resource that does not have a classical equivalent. The prevailing theory, however, posits that quantum entanglement represents merely a specific type of quantum correlation \cite{Henderson2001, Ollivier2001}, which is a more fundamental concept than quantum entanglement. The correlation between classical and quantum components may be more fundamental and extensive than entanglement. Ollivier and Zurek initially introduced the concept of quantum discord \cite{Ollivier2001, Zurek2003}, presenting it as a novel and compelling candidate for quantum correlation. Vedral et al. presented a comprehensive explanation of the concept \cite{Henderson2001, Vedral2003, Dakic2010}, which has recently been applied across various fields of study \cite{WangJieci2010, Fanchini2010, HuYaoHua2012, Mohamed2013, XieMeiQiu2013, FanKaiMing2013, LiRuiQi2014, Aldoshin2014, Mohamed2018, Mohamed2019, JiaZhihAhn2020, Radhakrishnan2020, Brown2021}. Quantum decoherence refers to the process through which quantum correlations diminish as a result of external environmental noise. The reality that the actual quantum system is not an entirely closed ideal system. The decoherence process is, therefore, unavoidable. The primary challenge lies in addressing decoherence, which fundamentally hinders the potential superiority of QIP.

This research significantly incorporates the quantum electrodynamics (QED) model, providing a distinctive physical framework for examining the interaction between light and matter. In this model, impurity two- or multi-level structures, often referred to as atoms, are linked to fields of cavities. The Jaynes--Cummings model (JCM) \cite{Jaynes1963} along with its generalizations and modifications \cite{Tavis1968, Angelakis2007} represents the most extensively studied cavity QED model and is typically more straightforward to implement in experimental settings. Cavity QED models have garnered significant attention in recent years \cite{WeiHuanhuan2021, Prasad2018, GuoLijuan2019, Smith2021, Dull2021, Afanasyev2021, Miao2023, Ozhigov2023, Miaohuihui2024, Li2024}. In cavity QED systems, quantum correlation has rarely been utilized in previous research. Some quantum computations may occasionally utilize quantum correlation. It is essential to examine the kinetics of quantum correlation in cavity QED systems.

We investigate quantum entanglement in bipartite systems, based on the JCM, by analyzing quantum entropy and concurrence. The quantum discord is acknowledged. Quantum discord is analyzed and compared with entanglement. The quantum master equation (QME) offers a solution for understanding the dissipative dynamics in open systems. We examine the effects of initial entanglement and dissipation intensity on quantum discord. We also examined the dynamics of quantum discord in the $\rm{OH}^+$ model, based on the Tavis--Cummings model (TCM).
	
\section{Quantum entanglement}
\label{sec:Entanglement}

\begin{figure}[t]
\centering
\includegraphics[width=0.45\textwidth]{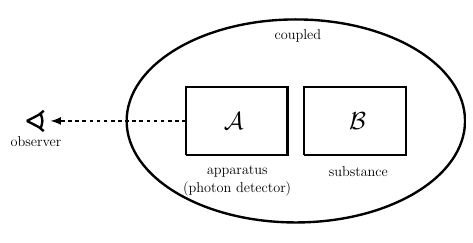}
\caption{(online color) {\it Coupling system.} The entire system is separated into two subsystems: the observation system $\mathcal{A}$ and the substance system $\mathcal{B}$. Subsystems are coupled together into a whole by the field in the optical cavity.}
\label{fig:CouplingSystem}
\end{figure}

Quantum entanglement can occur even when the particles are separated by significant distances. The issue of quantum entanglement, an essential aspect of quantum mechanics that is absent in classical mechanics, represents the fundamental difference between these two branches of physics. In a bipartite system, we designate the observed subsystem as $\mathcal{A}$ and the unobserved subsystem as $\mathcal{B}$ (see Fig. \ref{fig:CouplingSystem}).
	
\subsection{Quantum entropy}
\label{subsec:Entropy}
	
The von Neumann entropy of the reduced density matrix, utilized to assess quantum correlation \cite{Obada2007, Obada2008}, is applicable directly when the quantum system is closed and the entire system is in the pure state $|pure\rangle$. In this scenario, the von Neumann entropy of each of the two subsystems $\mathcal{A}$ and $\mathcal{B}$ can be utilized as an indicator of the system's entanglement, as established by the Schmidt decomposition \cite{Bennett1996}
\begin{equation}
	\label{eq:vonNeumannEntropy}
	E=S(\mathcal{A})=S(\mathcal{B})
\end{equation}
where $S(\mathcal{A}),\ S(\mathcal{B})$ are corresponding to $\rho_{\mathcal{A}},\ \rho_{\mathcal{B}}$, respectively. And $\rho_{\mathcal{A}},\ \rho_{\mathcal{B}}$ are reductions of density matrix $\rho_{\mathcal{AB}}$, which is the density matrix of whole system. $\rho_{\mathcal{A}},\ \rho_{\mathcal{B}}$ have the following forms
\begin{equation}
	\label{eq:ReductionRhoA}
	\rho_{\mathcal{A}}=Tr_{\mathcal{B}}(\rho_{\mathcal{AB}})=\sum_b(I_{\mathcal{A}}\otimes\langle b|_{\mathcal{B}})\rho_{\mathcal{AB}}(I_{\mathcal{A}}\otimes|b\rangle_{\mathcal{B}})
\end{equation}
\begin{equation}
	\label{eq:ReductionRhoB}
	\rho_{\mathcal{B}}=Tr_{\mathcal{A}}(\rho_{\mathcal{AB}})=\sum_a(\langle a|_{\mathcal{A}}\otimes I_{\mathcal{B}})\rho_{\mathcal{AB}}(|a\rangle_{\mathcal{A}}\otimes I_{\mathcal{B}})
\end{equation}
and
\begin{equation}
	\label{eq:DensityMatrix}
	\rho_{\mathcal{AB}}=|pure\rangle\langle pure|
\end{equation}
	
In quantum theory, the definition of von Neumann entropy of reduced density matrix $\rho_{\mathcal{A}(\mathcal{B})}$ is as follows
\begin{equation}
	\label{eq:vonNeumannTrace}
	S(\mathcal{A}(\mathcal{B}))=-Tr(\rho_{\mathcal{A}(\mathcal{B})} log_2\rho_{\mathcal{A}(\mathcal{B})})
\end{equation}
In terms of eigenvalues, we have
\begin{equation}
	\label{eq:vonNeumannEigen}
	S(\mathcal{A}(\mathcal{B}))=-\sum_i\lambda_i^{\mathcal{A}(\mathcal{B})}log_2\lambda_i^{\mathcal{A}(\mathcal{B})}
\end{equation}
where $\lambda_i^{\mathcal{A}(\mathcal{B})}$ --- eigenvalues of density matrix $\rho_{\mathcal{A}(\mathcal{B})}$. $0\leq S(\mathcal{A}(\mathcal{B}))\leq log_2N$, $N=2^n$, $N$ --- Hilbert space dimension, $n$ --- number of qubits. When $S(\mathcal{A}(\mathcal{B}))=0$, separable state is obtained, and $S(\mathcal{A}(\mathcal{B}))>0$ --- entangled state, especially, $S(\mathcal{A}(\mathcal{B}))=log_2N$ --- maximum entangled state.

\subsection{Concurrence}
\label{subsec:Concurrence}

We proceed to quantify the entanglement by examining concurrence within the two-qubit system, especially in the context of dissipation, where the entire system is characterized by a mixed state. Concurrence characterizes the entanglement of two qubits as
\begin{equation}
	\label{eq:Concurrence}
	\mathcal{C}=max\{\sqrt{\lambda_3}-\sqrt{\lambda_2}-\sqrt{\lambda_1}-\sqrt{\lambda_0},\ 0\}
\end{equation}
This equation is also called Hill--Wootters equation \cite{Hill1997, Wootters1998, Audenaert2001}, which is widely used in various aspects of physics, such as condensed matter physics \cite{Wang2002, Laflorencie2016, Lima2021}. Here $\lambda_1$, $\lambda_2$, $\lambda_3$ and $\lambda_4$ are eigenvalues of $\tilde{\rho}_{\mathcal{AB}}$, and $\lambda_3>\lambda_2>\lambda_1>\lambda_0\geq 0$. Density matrix $\tilde{\rho}_{\mathcal{AB}}$ has the following form
\begin{equation}
	\label{eq:TildeRho}
	\tilde{\rho}_{\mathcal{AB}}=\rho_{\mathcal{AB}}(\sigma_y\otimes\sigma_y)\rho_{\mathcal{AB}}^*(\sigma_y\otimes\sigma_y)^{\dag}
\end{equation}
where $\rho_{\mathcal{AB}}^*$ --- complex conjugation of a matrix $\rho_{\mathcal{AB}}$. $\sigma_y$ is defined as the component of the Pauli matrix in the $y$ direction while the eigenstates --- $|0\rangle$ and $|1\rangle$, and $\sigma_y=-i|0\rangle\langle 1|+i|1\rangle\langle 0|$. When $\mathcal{C}=0$, separable state is obtained, and $\mathcal{C}>0$ --- entangled state, especially, $\mathcal{C}=1$ --- maximum entangled state.

\section{Quantum discord}
\label{sec:Discord}
	
In quantum information theory, quantum discord serves as a metric for quantifying nonclassical correlations between two subsystems of a quantum system. It also includes correlations arising from quantum physical phenomena, though not exclusively from quantum entanglement. It serves as a measure for the quantumness of correlations.

\subsection{Overall correlation}
\label{subsec:Overall}
	
Initially, we present the principles of Shannon's theory within the domain of quantum informatics. In classical information theory, information can be quantified using Shannon entropy
\begin{equation}
	\label{eq:Shannon}
	H(X)=\sum_xP_{|X=x}lnP_{|X=x}
\end{equation}
where $P_{|X=x}$ represents the probability of $X$ taking the value $x$. For two random variables $X$ and $Y$, the total correlation between them can be quantified through the mutual information, which is defined as
\begin{equation}
	\label{eq:ClassicalMutual}
	I(X:Y)=H(X)+H(Y)-H(X,Y)
\end{equation}

We expand the aforementioned equation to encompass the domain of quantum information, resulting in the formulation of quantum mutual information
\begin{equation}
	\label{eq:QuantumMutual}
	\mathcal{I}(\mathcal{B}:\mathcal{A})=S(\mathcal{A})+S(\mathcal{B})-S(\mathcal{AB})=\mathcal{I}(\mathcal{A}:\mathcal{B})
\end{equation}
where $S(\mathcal{AB})$ --- von Neumann entropy for the system $\mathcal{AB}$ with density matrix $\rho_{\mathcal{AB}}$, $S(\mathcal{A})$ --- entropy for the subsystem $\mathcal{A}$ with density matrix $\rho_{\mathcal{A}}$, $S(\mathcal{B})$ --- entropy for the subsystem $\mathcal{B}$ with density matrix $\rho_{\mathcal{B}}$. $\mathcal{I}(\mathcal{B}:\mathcal{A})$ describes the overall correlation of the system.

\subsection{Classical correlation}
\label{subsec:Classical}
	
Subsequently, $\mathcal{J}(\mathcal{B}:\mathcal{A})$, representing the maximum classical correlation, is introduced. The following form is presented
\begin{equation}
	\label{eq:ClassicalCorrelationBA}
	\begin{aligned}
		\mathcal{J}(\mathcal{B}:\mathcal{A})&=\underset{\{\Pi^{\mathcal{A}}_k\}}{max}[S(\mathcal{B})-S(\mathcal{B}|\mathcal{A})]\\
		&=\underset{\{\Pi^{\mathcal{A}}_k\}}{max}[S(\mathcal{B})-\sum_kp_kS(\rho_k)]\\
		&=S(\mathcal{B})-\underset{\{\Pi^{\mathcal{A}}_k\}}{min}\sum_kp_kS(\rho_k)
	\end{aligned}
\end{equation}
where $\{\Pi_k^{\mathcal{A}}\}$ constitutes a complete set of projectors for the von Neumann projective measurement (orthogonal projection basis) of subsystem $\mathcal{A}$. The projective measurement basis for a two-dimensional Hilbert space is outlined as follows
\begin{equation}
	\label{eq:MeasurementvonNeumann1Qubits}
	\{|b_k\rangle\}=\{cos\theta|0\rangle+sin\theta|1\rangle,\ sin\theta|0\rangle-cos\theta|1\rangle\}
\end{equation}
where $\theta\in[0,\frac{\pi}{2}]$ and projection operators
\begin{subequations}
	\label{eq:Projecteurs}
	\begin{align}
		\Pi_0^{\mathcal{A}}&=cos^2\theta|0\rangle\langle0|+cos\theta sin\theta|0\rangle\langle1|\nonumber\\
		&+cos\theta sin\theta|1\rangle\langle0|+sin^2\theta|1\rangle\langle1|\label{eq:Projecteur0}\\
		\Pi_1^{\mathcal{A}}&=sin^2\theta|0\rangle\langle0|-cos\theta sin\theta|0\rangle\langle1|\nonumber\\
		&-cos\theta sin\theta|1\rangle\langle0|+cos^2\theta|1\rangle\langle1|\label{eq:Projecteur1}
	\end{align}
\end{subequations}
	
The von Neumann projective measurement basis in Eq. \eqref{eq:MeasurementvonNeumann1Qubits} can be extended to
\begin{equation}
	\label{eq:MeasurementGeneral}
	\{|b_k\rangle\}=\{cos\theta|0\rangle+sin\theta e^{i\varphi}|1\rangle,sin\theta e^{-i\varphi}|0\rangle-cos\theta|1\rangle\}
\end{equation}
where $\varphi\in[0,2\pi]$.
	
Following the construction of the projection operators for subsystem $\mathcal{A}$, measurements can be performed to obtain $\rho_k$, defined as follows
\begin{equation}
	\label{eq:RhokBA}
	\rho_k=Tr_{\mathcal{A}}[(\Pi_k^{\mathcal{A}}\otimes I_{\mathcal{B}})\rho_{\mathcal{AB}}(\Pi_k^{\mathcal{A}}\otimes I_{\mathcal{B}})^{\dag}]/p_k
\end{equation}
where
\begin{equation}
	\label{eq:pkBA}
	p_k=Tr[(\Pi_k^{\mathcal{A}}\otimes I_{\mathcal{B}})\rho_{\mathcal{AB}}(\Pi_k^{\mathcal{A}}\otimes I_{\mathcal{B}})^{\dag}]
\end{equation}
$\rho_k$ is the state of the subsystem $\mathcal{B}$ after a measurement of subsystem $\mathcal{A}$ leading to an outcome $k$ with a probability $p_k$.
	
It should be noted that in most cases $\mathcal{J}(\mathcal{B}:\mathcal{A})$ is not equal to $\mathcal{J}(\mathcal{A}:\mathcal{B})$, which is described as follows
\begin{equation}
	\label{eq:ClassicalCorrelationAB}
	\begin{aligned}
		\mathcal{J}(\mathcal{A}:\mathcal{B})&=\underset{\{\Pi^{\mathcal{B}}_{k'}\}}{max}[S(\mathcal{A})-S(\mathcal{A}|\mathcal{B})]\\
		&=\underset{\{\Pi^{\mathcal{B}}_{k'}\}}{max}[S(\mathcal{A})-\sum_{k'}p_{k'}S(\rho_{k'})]\\
		&=S(\mathcal{A})-\underset{\{\Pi^{\mathcal{B}}_{k'}\}}{min}\sum_{k'}p_{k'}S(\rho_{k'})
	\end{aligned}
\end{equation}
where
\begin{equation}
	\label{eq:RhokAB}
	\rho_{k'}=Tr_{\mathcal{B}}[(I_{\mathcal{A}}\otimes\Pi_{k'}^{\mathcal{B}})\rho_{\mathcal{AB}}(I_{\mathcal{A}}\otimes\Pi_{k'}^{\mathcal{B}})^{\dag}]/p_{k'}
\end{equation}
\begin{equation}
	\label{eq:pkAB}
	p_{k'}=Tr[(I_{\mathcal{A}}\otimes\Pi_{k'}^{\mathcal{B}})\rho_{\mathcal{AB}}(I_{\mathcal{A}}\otimes\Pi_{k'}^{\mathcal{B}})^{\dag}]
\end{equation}
	
\subsection{Quantum correlation}
\label{subsec:Quantum}

The quantum discord is defined as the difference between total correlation and the maximum classical correlation
\begin{equation}
	\label{eq:QuantumDiscord}
	\mathcal{D}(\mathcal{B}:\mathcal{A})=\mathcal{I}(\mathcal{B}:\mathcal{A})-\mathcal{J}(\mathcal{B}:\mathcal{A})
\end{equation}
where $0\leq \mathcal{D}(\mathcal{B}:\mathcal{A})<\mathcal{I}(\mathcal{B}:\mathcal{A}),\ \mathcal{D}(\mathcal{B}:\mathcal{A})\leq S(\mathcal{A})$. As in the case of the maximum classical correlation, $\mathcal{D}(\mathcal{B}:\mathcal{A})$ is not equal to $\mathcal{D}(\mathcal{A}:\mathcal{B})$ in most cases. The physical interpretation of quantum discord indicates that a higher value of $\mathcal{D}(\mathcal{B}:\mathcal{A})$ corresponds to a greater minimum loss of information in the subsystem $\mathcal{B}$ following the measurement of subsystem $\mathcal{A}$. The greater the minimum disturbance experienced by subsystem $\mathcal{B}$ following the measurement of subsystem $\mathcal{A}$, the stronger the correlation between $\mathcal{B}$ and $\mathcal{A}$. The maximization in $\mathcal{J}(\mathcal{B}:\mathcal{A})$ indicates the highest level of information acquired regarding the system $\mathcal{B}$ due to the implementation of perfect measurement $\{\Pi_k^{\mathcal{A}}\}$. The quantum discord is demonstrated to be zero for states exhibiting solely classical correlation, while it is nonzero for states that possess quantum correlation.

\section{Jaynes--Cummings model}
\label{sec:JCM}
	
This paper investigates the dynamics of quantum entanglement and quantum discord in two cavity QED models. Initially, the JCM is presented, detailing the interaction between a two-level atom and a quantized mode of an optical cavity. The JCM holds significant theoretical and experimental relevance in the fields of solid-state physics, QIP, atomic physics, and quantum optics.

\begin{figure*}[t]
\centering
\includegraphics[width=1.\textwidth]{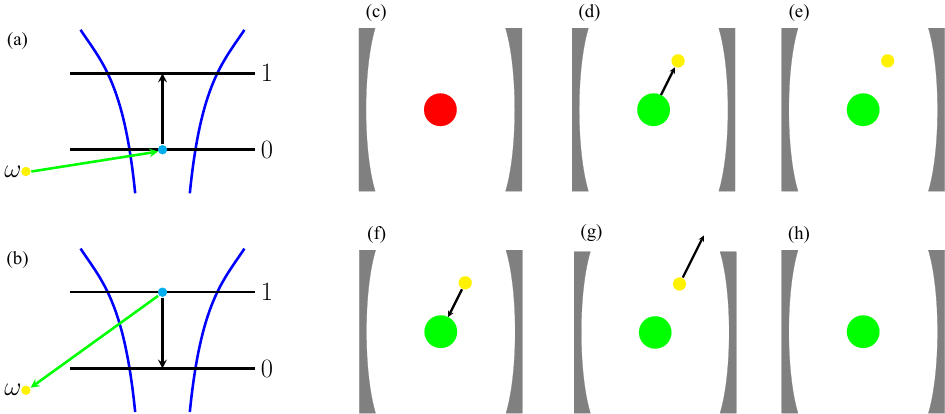}
\caption{(online color) {\it The JCM.} The yellow dot stands for photon. Electron is seen in panel (a) and panel (b) as blue dot. Excited atom and ground state atom are seen in panels (c) $\sim$ (h) as red and green dot, respectively.}
\label{fig:JCM}
\end{figure*}
	
The basic states of the JCM system are as follow
\begin{equation}
	\label{eq:BasisJCModel}
	|\Psi\rangle=\underbrace{|p\rangle_{ph}}_{\mathcal{A}}\underbrace{|l\rangle_{at}}_{\mathcal{B}}
\end{equation}
where $p=0$ means no free photons, $p=1$ means one free photon; $l=0$ --- electron in ground state, $l=1$ --- electron in excited state. The photon state is partly viewed as the observed by photon detector.
	
The interaction between the atom and the field is thoroughly detailed in Fig. \ref{fig:JCM}. Excitation takes place when an electron in the ground state, as depicted in panel (a), absorbs a photon and transitions to the excited state. De-excitation refers to the process in which an excited electron in panel (b) transitions to the ground state after the emission of a photon. Panels (c), (e), and (h) correspond respectively to three states for JCM: $|01\rangle,\ |10\rangle,\ |00\rangle$. Excitation is represented as (e) $\stackrel{\rm{(f)}}{\longrightarrow}$ (c). De-excitation is represented as (c) $\stackrel{\rm{(d)}}{\longrightarrow}$ (e). If the optical cavity is ideal, the system remains closed, and its state will oscillate between (c) and (e) due to the absence of photon leakage or dissipation. The system comprises the entangled states of $|01\rangle$ and $|10\rangle$, with the quantum discord presently being greater than or equal to zero. If the probability of either $|01\rangle$ or $|10\rangle$ is zero, then the quantum discord is zero. However, in practice, the optical cavity, referred to as an open system, is not flawless, and photons will escape from it into the surrounding environment. Dissipation is represented as (e) $\stackrel{\rm{(g)}}{\longrightarrow}$ (h). The atoms within the cavity will eventually reach stability in the ground state $|00\rangle$ as a result of photon leakage, indicating that the system will lose its quantum correlation and the quantum discord will become zero.
	
\subsection{Hamiltonian of JCM}
\label{subsec:HamilJCM}
	
Prior to constructing the Hamiltonian, we will first introduce the rotating wave approximation (RWA) \cite{Wu2007}, which will be considered
\begin{equation}
	\label{eq:RWACondition}
	\frac{g}{\hbar\omega_a}\approx\frac{g}{\hbar\omega_c}\ll 1
\end{equation}
For convenience, we assume that the electron transition frequency $\omega_a$ and field frequency $\omega_c$ are equal ($\omega=\omega_a=\omega_c$). And Hamiltonian of JCM has following form
\begin{equation}
	\label{eq:HamiltonianJCModel}
	H^{RWA}_{JCM}=\hbar\omega a^{\dag}a+\hbar\omega\varsigma^{\dag}\varsigma+g(a^{\dag}\varsigma+a\varsigma^{\dag})
\end{equation}
In this context, $\hbar$ represents the reduced Planck constant, while $g$ denotes the coupling strength between the photonic mode $\omega$ and the electron within the molecule. In this context, $a$ represents the photon annihilation operator, while $a^{\dag}$ denotes the photon creation operator. Additionally, $\varsigma$ signifies the electron relaxation operator, and $\varsigma^{\dag}$ indicates the electron excitation operator.
	
\subsection{Quantum master equation}
\label{subsec:QME}
	
The QME in the Markovian approximation for the density operator $\rho$ of the open system takes the following form
\begin{equation}
	\label{eq:QME}
	i\hbar\dot{\rho}=\left[H,\rho\right]+iL\left(\rho\right)
\end{equation}
where $\left[H,\rho\right]=H\rho-\rho H$ is the commutator. $L\left(\rho\right)$ is as follows
\begin{equation}
	\label{eq:LindbladOperator}
	L\left(\rho\right)=\sum_{k\in\mathcal{K}}\gamma_k\left(A_k\rho A_k^{\dag}-\frac{1}{2}\left\{\rho, A_k^{\dag}A_k\right\}\right)
\end{equation}
where $\mathcal{K}$ is the collection of all potential dissipations, $A_k$ is jump operator, and $\left\{\rho, A_k^{\dag}A_k\right\}={\rho A_k^{\dag}A_k + A_k^{\dag}A_k\rho}$ is the anticommutators. The term $\gamma_k$ refers to the overall spontaneous emission rate for photons for $k$-dissipation caused by photon leakage from the cavity to the external environment.	

\section{Tavis--Cummings model}
\label{sec:TCM}

\begin{figure*}[t]
\centering
\includegraphics[width=1.\textwidth]{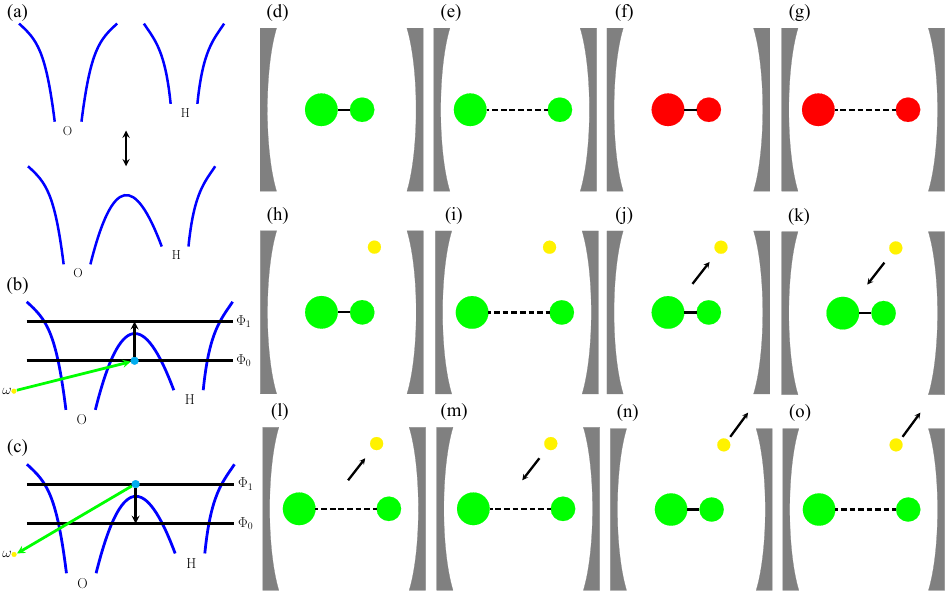}
\caption{(online color) {\it The $\rm{OH}^+$ model.} The large circle represents oxygen atom and the small one represents hydrogen atom. The solid line indicates a strong covalent bond, the dashed line indicates a weak covalent bond. The other symbols are the same as those in Fig. \ref{fig:JCM}.}
\label{fig:TCM}
\end{figure*}

The $\rm{OH}^+$ model is constructed within the framework of TCM. Two artificial two-level atoms are present in the optical cavity: an oxygen atom and a hydrogen atom. There exists a singular valence electron separating them. The hybridization and de-hybridization of orbitals are depicted in Fig. \ref{fig:TCM} (a). Figures \ref{fig:TCM} (b) and (c) illustrate the processes of excitation and de-excitation. The electron in the atomic ground orbital $0_{\rm{O}}$ (or $0_{\rm{H}}$) is bound to the nucleus and does not participate in the formation of chemical bonds. The subscripts O and H assigned to the eigenstates correspond to oxygen and hydrogen, respectively. Hybridization of atomic orbitals into molecular orbitals occurs exclusively when the electron occupies the atomic excited orbital $1_{\rm{O}}$ (or $1_{\rm{H}}$). This process results in the formation of the ground ${\Phi_0}_{mol}$ and excited ${\Phi_1}_{mol}$ molecular orbitals
\begin{equation}
	\label{eq:Phi0}
	{\Phi_0}_{mol}=\alpha 1_{\rm{O}}+\beta 1_{\rm{H}}
\end{equation}
\begin{equation}
	\label{eq:Phi1}
	{\Phi_1}_{mol}=\beta 1_{\rm{O}}-\alpha 1_{\rm{H}}
\end{equation}
where $\alpha$ and $\beta$ are positive coefficients depend on depth of potential wells in atoms, $\alpha>\beta,\ \alpha^2+\beta^2=1$.
	
Since there is only one valence electron in the model, the basic state is denoted as follows:
\begin{equation}
	\label{eq:BasisModelOH+}
	|\psi\rangle=\underbrace{|p\rangle_{ph}}_{\mathcal{A}}\underbrace{|l\rangle_{mol}|k\rangle_n}_{\mathcal{B}}
\end{equation}
where $p$ is number of photons, $n=0,1$; $l=0$ means electron in molecular ground orbital ${\Phi_0}_{mol}$, $l=1$ means electron in molecular excited orbital ${\Phi_1}_{mol}$; $k=0$ means the nuclei are close from each other (at this point covalent bond is strong), and $k=1$ means far away (at this point covalent bond is weak).
	
The interactions and dissipations of the $\rm{OH}^+$ model are detailed in Fig. \ref{fig:TCM}. Panels (d) $\sim$ (i) correspond to six states for the $\rm{OH}^+$ model: $|000\rangle$, $|001\rangle$, $|010\rangle$, $|011\rangle$, $|100\rangle$, $|101\rangle$. Excitations are shown as (h) $\stackrel{\rm{(k)}}{\longrightarrow}$ (f), (i) $\stackrel{\rm{(m)}}{\longrightarrow}$ (g); de-excitations are shown as (f) $\stackrel{\rm{(j)}}{\longrightarrow}$ (h), (g) $\stackrel{\rm{(l)}}{\longrightarrow}$ (i); dissipations are shown as (h) $\stackrel{\rm{(n)}}{\longrightarrow}$ (d), (i) $\stackrel{\rm{(o)}}{\longrightarrow}$ (e).
	
\subsection{Hamiltonian of the $\rm{OH}^+$ model}
\label{subsec:HamilTCM}
	
Hamiltonian of the $\rm{OH}^+$ model has following form
\begin{equation}
	\label{eq:HamiltonianModelOH+}
	H^{RWA}_{\rm{OH}^+}=H_0+H_{int}
\end{equation}
where
\begin{equation}
	\label{eq:HamiltonianModelOH+0}
	H_0=\hbar\omega_c a^\dagger a+\hbar\omega_b\varsigma_b^\dagger\varsigma_b+\hbar\omega_a\varsigma_a^\dagger\varsigma_a
\end{equation}
\begin{equation}
	\label{eq:HamiltonianModelOH+Int}
	H_{int}=g_b(\varsigma_b+\varsigma_b^\dagger)+g_a(a^\dagger\varsigma_a+a\varsigma_a^\dagger)
\end{equation}
where $a,\ a^{\dag}$ are photon annihilation and creation operators, respectively; $\varsigma_b,\ \varsigma_b^{\dag}$ are covalent bond breaking and formation operator, respectively; $\varsigma_a,\ \varsigma_a^{\dag}$ are electron relaxation and excitation operator, respectively; $\omega_c$ is photon frequency in the cavity; $\omega_b$ is phonon frequency, corresponding to the energy of covalent bond; $\omega_a$ is atomic transition frequency, and $\omega_c=\omega_a=\omega$; $g_b$ is strength of formation of covalent bond; $g_a$ is strength of interaction of a molecule with the field.
	
The system exhibits greater instability when the electron is in an excited state compared to its ground state. This suggests that covalent bonds can be formed or disrupted with relative ease. Currently, $g_b=g_{b_1}$, which represents a significant value. The system exhibits increased stability when the electron occupies the ground state. Covalent bonds will be challenging to form or dissociate. $g_b=g_{b_0}$, which is currently weak, and $g_{b_1}\gg g_{b_0}$.

\begin{figure}[t]
\centering
\includegraphics[width=0.5\textwidth]{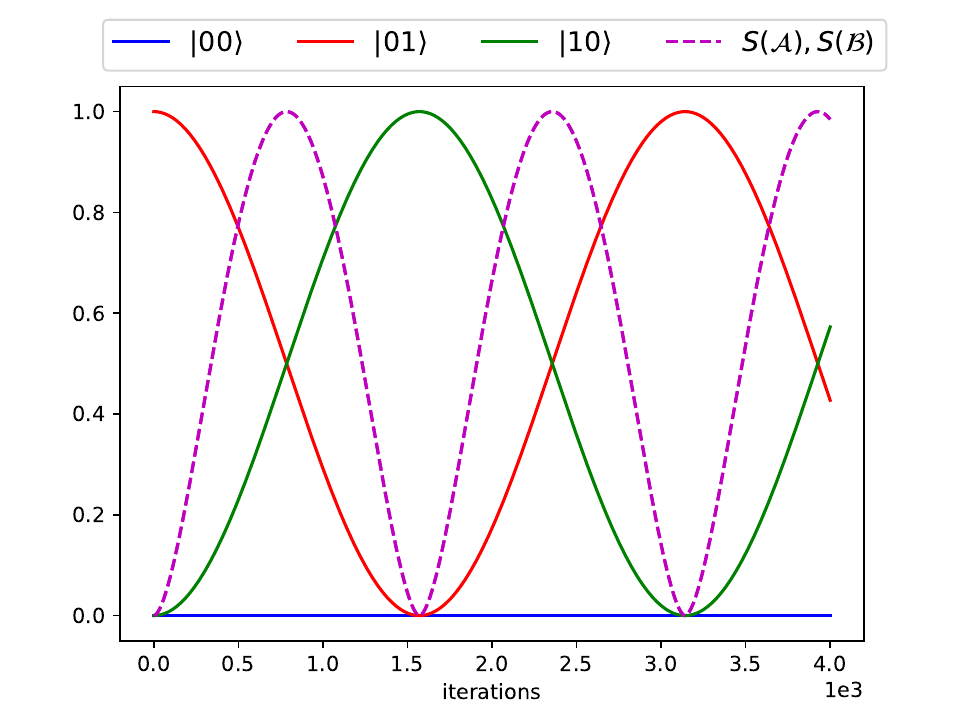}
\caption{(online color) {\it Quantum evolution of closed system.} Here $\alpha=0,\ \gamma=0$. Blue, green and red solid curves represent the probability of states. Magenta dashed curve --- entropy.}
\label{fig:EvolutionJCMClosed}
\end{figure}

Additionally, it is important to note that a strong covalent bond is formed when two atoms are in close proximity, suggesting that the diatomic system is approaching the molecular configuration; at this point, $g_a=g_{a_0}$, which is substantial. Conversely, a weak covalent bond arises when two atoms are distanced from each other, indicating that the diatomic system is leaning towards the independent atomic state; at this juncture, $g_a=g_{a_1}$, which is minimal. We assume that $g_{a_1}\ll g_{a_0}$.

The QME of the $\rm{OH}^+$ model exhibits similarities to the discussion presented in the previous subsection.

\section{Results}
\label{sec:Results}
	
This paper presents a schematic model of quantum evolutions achieved through the resolution of the QME. The QME approach has been utilized to investigate the dynamics of quantum open systems \cite{Breuer2002}, and it aligns with the principles of quantum thermodynamics \cite{Alicki1979, Kosloff2013}. This is applicable solely for Markovian approximation. The solution $\rho(t)$ in Eq. \eqref{eq:QME} can be approximated through a two-step process: initially, we take a step in solving the unitary component of Eq. \eqref{eq:QME}
\begin{equation}
	\label{eq:UnitaryPart}
	\tilde{\rho}(t+dt)=exp({-\frac{i}{\hbar}Hdt})\rho(t)exp(\frac{i}{\hbar}Hdt)
\end{equation}
and in the second step, make one step in the solution of Eq. \eqref{eq:QME} with the commutator removed
\begin{equation}
	\label{eq:Solution}
	\rho(t+dt)=\tilde{\rho}(t+dt)+\frac{1}{\hbar}L(\tilde{\rho}(t+dt))dt
\end{equation}
In each iteration, an updated $\rho(t+dt)$ is obtained and subsequently utilized in the calculations of entropy, concurrence, and quantum discord, yielding time-dependent results.
	
\subsection{Two-qubit system}
\label{subsec:TwoQubit}

\begin{figure*}[t]
\centering
\includegraphics[width=0.7\textwidth]{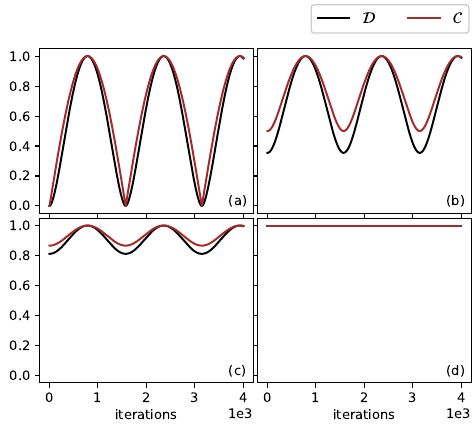}
\caption{(online color) {\it Quantum correlation dynamics in closed JCM with different initial entanglements.} $\gamma=0$ in these cases. Here (a) $\alpha=0$, (b) $\alpha=\frac{\pi}{12}$, (c) $\alpha=\frac{\pi}{6}$, (d) $\alpha=\frac{\pi}{4}$. Quantum discord is represented by black curve and concurrence --- red curve.}
\label{fig:InitialEntanglements}
\end{figure*}
	
In simulations for two-qubit system: $\hbar=1$, $\omega=10^8$, $g=10^6$. The initial state is defined as
\begin{equation}
	\label{eq:InitialJCModel}
	|\Psi(0)\rangle=cos\alpha|01\rangle+sin\alpha|10\rangle
\end{equation}
where $\alpha\in[0,\frac{\pi}{4}]$.

In Fig. \ref{fig:EvolutionJCMClosed}, we suppose the initial state $|\Psi(0)\rangle=|01\rangle$ ($\alpha=0$), and $\gamma=0$. When the system is closed, the quantum evolution presents Rabi oscillation \cite{Rabi1936} of $|01\rangle$ and $|10\rangle$ in Fig. \ref{fig:EvolutionJCMClosed}. At this time, the $|00\rangle$ is always $0$, because there is no photon leakage. We set the state of entire system as
\begin{equation}
	\label{eq:Fig4}
	|\Psi\rangle=a|01\rangle+b|10\rangle
\end{equation}
where $a^2+b^2=1$. For a bipartite system, $S(\mathcal{A})$ is always equal to $S(\mathcal{B})$. When coefficients $a$ and $b$ from Eq. \eqref{eq:Fig4} are equal, and $|\Psi\rangle=\frac{1}{\sqrt{2}}|01\rangle+\frac{1}{\sqrt{2}}|10\rangle$. $S(\mathcal{A})$ ($S(\mathcal{B})$) is equal to $1$ at this time ($S_{max}(\mathcal{A})=log_2N$, where $N=2^n=2$, because subsystem consists of a qubit ($n=1$), thus $S_{max}(\mathcal{A})=1$), and the maximum entanglement is obtained. When $a=1,\ b=0$ (or $a=0,\ b=1$), the entanglement of system is $0$ ($S(\mathcal{A})=S(\mathcal{B})=0$).
	
In Fig. \ref{fig:InitialEntanglements}, the measurement of quantum entanglement is conducted via the introduction of concurrence. Quantum discord is introduced as a means to quantify quantum correlation. In panel (a), both concurrence and quantum discord, similar to entropy in Fig. \ref{fig:EvolutionJCMClosed}, reach their maximum values of $1$ when the system exhibits the highest degree of entanglement and $0$ when quantum entanglement is absent. Additionally, von Neumann entropy is completely aligned with quantum discord and concurrence. A comparison of four panels indicates that the lower limit of quantum discord and concurrence increases with a higher initial degree of entanglement ($\alpha$). The initial degree of entanglement is $0$, and the lower bound remains $0$. In other terms, the quantum discord and concurrence curves exhibit fluctuations between $0$ and $1$. The curves progressively converge toward the $1$ boundary as $\alpha$ rises. The initial entanglement reaches its peak when $\alpha$ is $\frac{\pi}{4}$, at which stage the curves transform into horizontal lines, consistently maintaining a value of $1$.

\begin{figure}[t]
\centering
\includegraphics[width=0.5\textwidth]{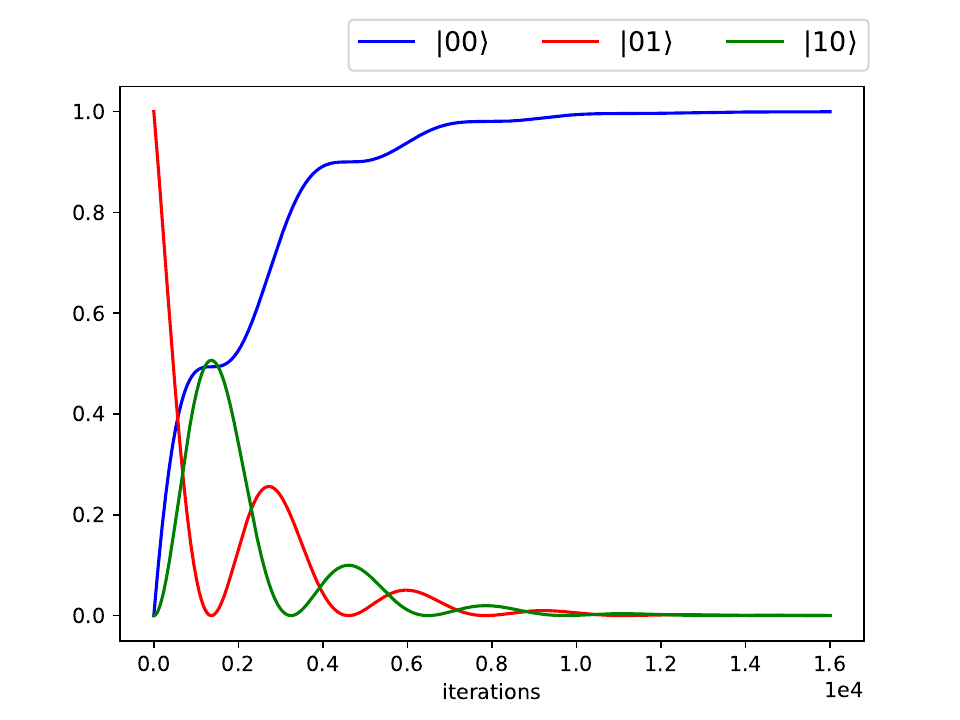}
\caption{(online color) {\it Quantum evolution of open system.} Here $\alpha=0,\ \gamma=g$. Curve definitions are the same as Fig. \ref{fig:EvolutionJCMClosed}.}
\label{fig:EvolutionJCMOpen}
\end{figure}

\begin{figure*}[t]
\centering
\includegraphics[width=0.7\textwidth]{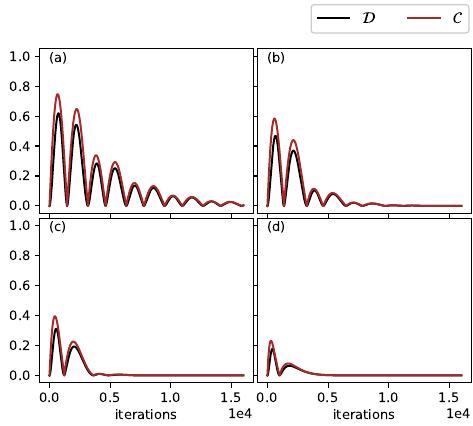}
\caption{(online color) {\it Quantum correlation dynamics in open JCM with different dissipation strengths.} $\alpha=0$ in these cases. Here (a) $\gamma=0.5g$, (b) $\gamma=g$, (c) $\gamma=2g$, (d) $\gamma=4g$.}
\label{fig:DissipationStrengths}
\end{figure*}

We will now concentrate on the open quantum system where $\gamma=g$. The entire system is currently not in a pure state. In Fig. \ref{fig:EvolutionJCMOpen}, the presence of photon leakage results in the absence of oscillation between the $|01\rangle$ and $|10\rangle$ states. The likelihood of $|00\rangle$ steadily rises from $0$ to $1$. We will now establish the state of the entire system as
\begin{equation}
	\label{eq:Fig6a}
	|\Psi\rangle=a|01\rangle+b|10\rangle+c|00\rangle
\end{equation}
where $a^2+b^2+c^2=1$. According to Fig. \ref{fig:EvolutionJCMOpen} (a), the probability of $|00\rangle$ is always non-zero when the evolution begins. And $a|01\rangle + c|00\rangle=|0\rangle(a|1\rangle+c|0\rangle)$, which shows that it is not entangled. Similarly, $b|10\rangle + c|00\rangle=(b|1\rangle+c|0\rangle)|0\rangle$, which are also not entangled. So the entanglement property of the whole system is still determined by $|01\rangle$ and $|10\rangle$.
	
In the event of photon leakage, the quantum correlation of the system will gradually diminish to $0$ (at which point, the system exists solely in the state of $|00\rangle$), as illustrated by the quantum discord in Fig. \ref{fig:DissipationStrengths}. The comparison of the panels in Fig. \ref{fig:DissipationStrengths} indicates that the oscillation of quantum discord and concurrence diminishes more rapidly as $\gamma$ increases. The oscillation occurs exclusively when $\gamma$ is $0$ (refer to Fig. \ref{fig:InitialEntanglements}).
	
\subsection{Three-qubit system}
\label{subsec:ThreeQubit}

\begin{figure*}[t]
\centering
\includegraphics[width=0.7\textwidth]{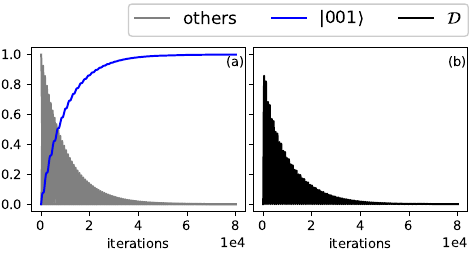}
\caption{(online color) {\it Quantum evolution and quantum discord dynamics in $\rm{OH}^+$ model.} $|001\rangle$ is represented by blue curve. Quantum evolution is shown in panel (a), and quantum discord dynamics is represented in panel (b).}
\label{fig:EvolutionCorrelationTCM}
\end{figure*}	
	
In simulations for three-qubit system: $\hbar=1$, $\omega_c=\omega_a=10^9$, $\omega_b=10^8$, $g_{b_0}=10^4,\ g_{b_1}=100g_{b_0},\ g_{a_1}=2g_{b_1},\ g_{a_0}=100g_{a_1}$. $g_{a_0}$, $g_{a_1}$, $g_{b_0}$, and $g_{b_1}$ are constants due to the quantum motion of nuclei. The tunneling phenomenon, characterized by the instantaneous movement of two atomic nuclei from a distant location to a nearby one, is known as the quantum mobility of nuclei. The strengths of covalent bond creation or breaking are constants at this moment. The initial state is defined as follows
\begin{equation}
	\label{eq:InitialModelOH+}
	|\Psi(0)\rangle=|101\rangle
\end{equation}
In the initial state, the atoms are positioned sufficiently far apart to neglect mutual interactions, a photon occupies the cavity, and an electron resides at the ground level.

The quantum evolution of the $\rm{OH}^+$ model, with a focus on photon dissipation related exclusively to electron transition, is illustrated in panel (a) of Fig. \ref{fig:EvolutionCorrelationTCM}. The blue curve, corresponding to the state $|001\rangle$, increases to $1$ throughout the evolution, whereas the other states quickly diminish to $0$. Furthermore, the quantum discord dynamics of this system are illustrated in panel (b). In violent oscillations, the quantum discord initially almost reaches $1$ and then gradually decays to $0$ (at this stage, only the state $|001\rangle$ remains in the system). This indicates that the quantum correlation between subsystems $\mathcal{A}$ and $\mathcal{B}$ progressively diminishes to $0$.

\section{Conclusion}
\label{sec:Conclusion}
	
This paper presents a simulation of the dynamics of quantum entanglement and quantum discord within cavity QED models. We have obtained several analytical results from it:

Initially, we analyze the two-qubit JCM system. Our findings indicate that within the closed system, quantum discord is aligned with entanglement. Furthermore, it has been observed that the lower limit of quantum discord and concurrence rises in accordance with the level of original entanglement; ideally, quantum discord and concurrence stabilize as a horizontal line when the initial entanglement attains its peak. Quantum discord is subsequently aligned with quantum entanglement (concurrence) within the open system. Simultaneously, it is observed that quantum discord and concurrence diminish more rapidly as the intensity of dissipation increases. Secondly, we examine the three-qubit TCM system and observe that the quantum discord experiences a gradual decay to $0$ following an initial period of significant fluctuation.

Although we utilized bipartite measurement to investigate the dynamics of quantum entanglement and quantum discord in fundamental cavity QED models, the results we achieved will provide a basis for future exploration of more complex variations of the cavity QED model.

\section*{Acknowledgments}
The reported study was funded by China Scholarship Council, project numbers 202108090483, 202108090327.

\bibliography{bibliography}

\begin{thebibliography}{52}%
\makeatletter
\providecommand \@ifxundefined [1]{%
 \@ifx{#1\undefined}
}%
\providecommand \@ifnum [1]{%
 \ifnum #1\expandafter \@firstoftwo
 \else \expandafter \@secondoftwo
 \fi
}%
\providecommand \@ifx [1]{%
 \ifx #1\expandafter \@firstoftwo
 \else \expandafter \@secondoftwo
 \fi
}%
\providecommand \natexlab [1]{#1}%
\providecommand \enquote  [1]{``#1''}%
\providecommand \bibnamefont  [1]{#1}%
\providecommand \bibfnamefont [1]{#1}%
\providecommand \citenamefont [1]{#1}%
\providecommand \href@noop [0]{\@secondoftwo}%
\providecommand \href [0]{\begingroup \@sanitize@url \@href}%
\providecommand \@href[1]{\@@startlink{#1}\@@href}%
\providecommand \@@href[1]{\endgroup#1\@@endlink}%
\providecommand \@sanitize@url [0]{\catcode `\\12\catcode `\$12\catcode
  `\&12\catcode `\#12\catcode `\^12\catcode `\_12\catcode `\%12\relax}%
\providecommand \@@startlink[1]{}%
\providecommand \@@endlink[0]{}%
\providecommand \url  [0]{\begingroup\@sanitize@url \@url }%
\providecommand \@url [1]{\endgroup\@href {#1}{\urlprefix }}%
\providecommand \urlprefix  [0]{URL }%
\providecommand \Eprint [0]{\href }%
\providecommand \doibase [0]{https://doi.org/}%
\providecommand \selectlanguage [0]{\@gobble}%
\providecommand \bibinfo  [0]{\@secondoftwo}%
\providecommand \bibfield  [0]{\@secondoftwo}%
\providecommand \translation [1]{[#1]}%
\providecommand \BibitemOpen [0]{}%
\providecommand \bibitemStop [0]{}%
\providecommand \bibitemNoStop [0]{.\EOS\space}%
\providecommand \EOS [0]{\spacefactor3000\relax}%
\providecommand \BibitemShut  [1]{\csname bibitem#1\endcsname}%
\let\auto@bib@innerbib\@empty
\bibitem [{\citenamefont {Horodecki}\ \emph {et~al.}(2009)\citenamefont
  {Horodecki}, \citenamefont {Horodecki}, \citenamefont {Horodecki},\ and\
  \citenamefont {Horodecki}}]{Horodecki2009}%
  \BibitemOpen
  \bibfield  {author} {\bibinfo {author} {\bibfnamefont {R.}~\bibnamefont
  {Horodecki}}, \bibinfo {author} {\bibfnamefont {P.}~\bibnamefont
  {Horodecki}}, \bibinfo {author} {\bibfnamefont {M.}~\bibnamefont
  {Horodecki}},\ and\ \bibinfo {author} {\bibfnamefont {K.}~\bibnamefont
  {Horodecki}},\ }\bibfield  {title} {\bibinfo {title} {Quantum entanglement},\
  }\href {https://doi.org/10.1103/RevModPhys.81.865} {\bibfield  {journal}
  {\bibinfo  {journal} {Rev. Mod. Phys.}\ }\textbf {\bibinfo {volume} {81}},\
  \bibinfo {pages} {865} (\bibinfo {year} {2009})}\BibitemShut {NoStop}%
\bibitem [{\citenamefont {Einstein}\ \emph {et~al.}(1935)\citenamefont
  {Einstein}, \citenamefont {Podolsky},\ and\ \citenamefont
  {Rosen}}]{Einstein1935}%
  \BibitemOpen
  \bibfield  {author} {\bibinfo {author} {\bibfnamefont {A.}~\bibnamefont
  {Einstein}}, \bibinfo {author} {\bibfnamefont {B.}~\bibnamefont {Podolsky}},\
  and\ \bibinfo {author} {\bibfnamefont {N.}~\bibnamefont {Rosen}},\ }\bibfield
   {title} {\bibinfo {title} {Can quantum-mechanical description of physical
  reality be considered complete?},\ }\href
  {https://doi.org/10.1103/PhysRev.47.777} {\bibfield  {journal} {\bibinfo
  {journal} {Phys. Rev.}\ }\textbf {\bibinfo {volume} {47}},\ \bibinfo {pages}
  {777} (\bibinfo {year} {1935})}\BibitemShut {NoStop}%
\bibitem [{\citenamefont {Nielsen}\ and\ \citenamefont
  {Chuang}(2010)}]{Nielsen2010}%
  \BibitemOpen
  \bibfield  {author} {\bibinfo {author} {\bibfnamefont {M.}~\bibnamefont
  {Nielsen}}\ and\ \bibinfo {author} {\bibfnamefont {I.}~\bibnamefont
  {Chuang}},\ }in\ \href {https://doi.org/10.1017/CBO9780511976667} {\emph
  {\bibinfo {booktitle} {Quantum Computation and Quantum Information: 10th
  Anniversary Edition}}}\ (\bibinfo  {publisher} {Cambridge University Press},\
  \bibinfo {address} {Cambridge},\ \bibinfo {year} {2010})\BibitemShut
  {NoStop}%
\bibitem [{\citenamefont {Ekert}(1991)}]{Ekert1991}%
  \BibitemOpen
  \bibfield  {author} {\bibinfo {author} {\bibfnamefont {A.~K.}\ \bibnamefont
  {Ekert}},\ }\bibfield  {title} {\bibinfo {title} {Quantum cryptography based
  on {B}ell's theorem},\ }\href {https://doi.org/10.1103/PhysRevLett.67.661}
  {\bibfield  {journal} {\bibinfo  {journal} {Phys. Rev. Lett.}\ }\textbf
  {\bibinfo {volume} {67}},\ \bibinfo {pages} {661} (\bibinfo {year}
  {1991})}\BibitemShut {NoStop}%
\bibitem [{\citenamefont {Bennett}\ and\ \citenamefont
  {Wiesner}(1992)}]{Bennett1992}%
  \BibitemOpen
  \bibfield  {author} {\bibinfo {author} {\bibfnamefont {C.~H.}\ \bibnamefont
  {Bennett}}\ and\ \bibinfo {author} {\bibfnamefont {S.~J.}\ \bibnamefont
  {Wiesner}},\ }\bibfield  {title} {\bibinfo {title} {Communication via one-
  and two-particle operators on {E}instein--{P}odolsky--{R}osen states},\
  }\href {https://doi.org/10.1103/PhysRevLett.69.2881} {\bibfield  {journal}
  {\bibinfo  {journal} {Phys. Rev. Lett.}\ }\textbf {\bibinfo {volume} {69}},\
  \bibinfo {pages} {2881} (\bibinfo {year} {1992})}\BibitemShut {NoStop}%
\bibitem [{\citenamefont {Bennett}\ \emph {et~al.}(1993)\citenamefont
  {Bennett}, \citenamefont {Brassard}, \citenamefont {Cr\'epeau}, \citenamefont
  {Jozsa}, \citenamefont {Peres},\ and\ \citenamefont
  {Wootters}}]{Bennett1993}%
  \BibitemOpen
  \bibfield  {author} {\bibinfo {author} {\bibfnamefont {C.~H.}\ \bibnamefont
  {Bennett}}, \bibinfo {author} {\bibfnamefont {G.}~\bibnamefont {Brassard}},
  \bibinfo {author} {\bibfnamefont {C.}~\bibnamefont {Cr\'epeau}}, \bibinfo
  {author} {\bibfnamefont {R.}~\bibnamefont {Jozsa}}, \bibinfo {author}
  {\bibfnamefont {A.}~\bibnamefont {Peres}},\ and\ \bibinfo {author}
  {\bibfnamefont {W.~K.}\ \bibnamefont {Wootters}},\ }\bibfield  {title}
  {\bibinfo {title} {Teleporting an unknown quantum state via dual classical
  and {E}instein--{P}odolsky--{R}osen channels},\ }\href
  {https://doi.org/10.1103/PhysRevLett.70.1895} {\bibfield  {journal} {\bibinfo
   {journal} {Phys. Rev. Lett.}\ }\textbf {\bibinfo {volume} {70}},\ \bibinfo
  {pages} {1895} (\bibinfo {year} {1993})}\BibitemShut {NoStop}%
\bibitem [{\citenamefont {Miao}(2024)}]{Miao2024}%
  \BibitemOpen
  \bibfield  {author} {\bibinfo {author} {\bibfnamefont {H.-h.}\ \bibnamefont
  {Miao}},\ }\bibfield  {title} {\bibinfo {title} {Investigating entropic
  dynamics of multiqubit cavity qed system},\ }\href
  {https://doi.org/https://doi.org/10.1002/qute.202400246} {\bibfield
  {journal} {\bibinfo  {journal} {Advanced Quantum Technologies}\ ,\ \bibinfo
  {pages} {2400246}} (\bibinfo {year} {2024})}\BibitemShut {NoStop}%
\bibitem [{\citenamefont {Henderson}\ and\ \citenamefont
  {Vedral}(2001)}]{Henderson2001}%
  \BibitemOpen
  \bibfield  {author} {\bibinfo {author} {\bibfnamefont {L.}~\bibnamefont
  {Henderson}}\ and\ \bibinfo {author} {\bibfnamefont {V.}~\bibnamefont
  {Vedral}},\ }\bibfield  {title} {\bibinfo {title} {Classical, quantum and
  total correlations},\ }\href {https://doi.org/10.1088/0305-4470/34/35/315}
  {\bibfield  {journal} {\bibinfo  {journal} {Journal of Physics A:
  Mathematical and General}\ }\textbf {\bibinfo {volume} {34}},\ \bibinfo
  {pages} {6899} (\bibinfo {year} {2001})}\BibitemShut {NoStop}%
\bibitem [{\citenamefont {Ollivier}\ and\ \citenamefont
  {Zurek}(2001)}]{Ollivier2001}%
  \BibitemOpen
  \bibfield  {author} {\bibinfo {author} {\bibfnamefont {H.}~\bibnamefont
  {Ollivier}}\ and\ \bibinfo {author} {\bibfnamefont {W.~H.}\ \bibnamefont
  {Zurek}},\ }\bibfield  {title} {\bibinfo {title} {Quantum discord: A measure
  of the quantumness of correlations},\ }\href
  {https://doi.org/10.1103/PhysRevLett.88.017901} {\bibfield  {journal}
  {\bibinfo  {journal} {Phys. Rev. Lett.}\ }\textbf {\bibinfo {volume} {88}},\
  \bibinfo {pages} {017901} (\bibinfo {year} {2001})}\BibitemShut {NoStop}%
\bibitem [{\citenamefont {Zurek}(2003)}]{Zurek2003}%
  \BibitemOpen
  \bibfield  {author} {\bibinfo {author} {\bibfnamefont {W.~H.}\ \bibnamefont
  {Zurek}},\ }\bibfield  {title} {\bibinfo {title} {Quantum discord and
  {M}axwell's demons},\ }\href {https://doi.org/10.1103/PhysRevA.67.012320}
  {\bibfield  {journal} {\bibinfo  {journal} {Phys. Rev. A}\ }\textbf {\bibinfo
  {volume} {67}},\ \bibinfo {pages} {012320} (\bibinfo {year}
  {2003})}\BibitemShut {NoStop}%
\bibitem [{\citenamefont {Vedral}(2003)}]{Vedral2003}%
  \BibitemOpen
  \bibfield  {author} {\bibinfo {author} {\bibfnamefont {V.}~\bibnamefont
  {Vedral}},\ }\bibfield  {title} {\bibinfo {title} {Classical correlations and
  entanglement in quantum measurements},\ }\href
  {https://doi.org/10.1103/PhysRevLett.90.050401} {\bibfield  {journal}
  {\bibinfo  {journal} {Phys. Rev. Lett.}\ }\textbf {\bibinfo {volume} {90}},\
  \bibinfo {pages} {050401} (\bibinfo {year} {2003})}\BibitemShut {NoStop}%
\bibitem [{\citenamefont {Daki\ifmmode~\acute{c}\else \'{c}\fi{}}\ \emph
  {et~al.}(2010)\citenamefont {Daki\ifmmode~\acute{c}\else \'{c}\fi{}},
  \citenamefont {Vedral},\ and\ \citenamefont {Brukner}}]{Dakic2010}%
  \BibitemOpen
  \bibfield  {author} {\bibinfo {author} {\bibfnamefont {B.}~\bibnamefont
  {Daki\ifmmode~\acute{c}\else \'{c}\fi{}}}, \bibinfo {author} {\bibfnamefont
  {V.}~\bibnamefont {Vedral}},\ and\ \bibinfo {author} {\bibfnamefont
  {i.~c.~v.}\ \bibnamefont {Brukner}},\ }\bibfield  {title} {\bibinfo {title}
  {Necessary and sufficient condition for nonzero quantum discord},\ }\href
  {https://doi.org/10.1103/PhysRevLett.105.190502} {\bibfield  {journal}
  {\bibinfo  {journal} {Phys. Rev. Lett.}\ }\textbf {\bibinfo {volume} {105}},\
  \bibinfo {pages} {190502} (\bibinfo {year} {2010})}\BibitemShut {NoStop}%
\bibitem [{\citenamefont {Wang}\ \emph {et~al.}(2010)\citenamefont {Wang},
  \citenamefont {Deng},\ and\ \citenamefont {Jing}}]{WangJieci2010}%
  \BibitemOpen
  \bibfield  {author} {\bibinfo {author} {\bibfnamefont {J.}~\bibnamefont
  {Wang}}, \bibinfo {author} {\bibfnamefont {J.}~\bibnamefont {Deng}},\ and\
  \bibinfo {author} {\bibfnamefont {J.}~\bibnamefont {Jing}},\ }\bibfield
  {title} {\bibinfo {title} {Classical correlation and quantum discord sharing
  of {D}irac fields in noninertial frames},\ }\href
  {https://doi.org/10.1103/PhysRevA.81.052120} {\bibfield  {journal} {\bibinfo
  {journal} {Phys. Rev. A}\ }\textbf {\bibinfo {volume} {81}},\ \bibinfo
  {pages} {052120} (\bibinfo {year} {2010})}\BibitemShut {NoStop}%
\bibitem [{\citenamefont {Fanchini}\ \emph {et~al.}(2010)\citenamefont
  {Fanchini}, \citenamefont {Castelano},\ and\ \citenamefont
  {Caldeira}}]{Fanchini2010}%
  \BibitemOpen
  \bibfield  {author} {\bibinfo {author} {\bibfnamefont {F.~F.}\ \bibnamefont
  {Fanchini}}, \bibinfo {author} {\bibfnamefont {L.~K.}\ \bibnamefont
  {Castelano}},\ and\ \bibinfo {author} {\bibfnamefont {A.~O.}\ \bibnamefont
  {Caldeira}},\ }\bibfield  {title} {\bibinfo {title} {Entanglement versus
  quantum discord in two coupled double quantum dots},\ }\href
  {https://doi.org/10.1088/1367-2630/12/7/073009} {\bibfield  {journal}
  {\bibinfo  {journal} {New Journal of Physics}\ }\textbf {\bibinfo {volume}
  {12}},\ \bibinfo {pages} {073009} (\bibinfo {year} {2010})}\BibitemShut
  {NoStop}%
\bibitem [{\citenamefont {Hu}\ and\ \citenamefont {Fang}(2012)}]{HuYaoHua2012}%
  \BibitemOpen
  \bibfield  {author} {\bibinfo {author} {\bibfnamefont {Y.-H.}\ \bibnamefont
  {Hu}}\ and\ \bibinfo {author} {\bibfnamefont {M.-F.}\ \bibnamefont {Fang}},\
  }\bibfield  {title} {\bibinfo {title} {Quantum discord between two moving
  two-level atoms},\ }\href {https://doi.org/doi:10.2478/s11534-011-0076-6}
  {\bibfield  {journal} {\bibinfo  {journal} {Open Physics}\ }\textbf {\bibinfo
  {volume} {10}},\ \bibinfo {pages} {145} (\bibinfo {year} {2012})}\BibitemShut
  {NoStop}%
\bibitem [{\citenamefont {Mohamed}(2013)}]{Mohamed2013}%
  \BibitemOpen
  \bibfield  {author} {\bibinfo {author} {\bibfnamefont {A.-B.~A.}\
  \bibnamefont {Mohamed}},\ }\bibfield  {title} {\bibinfo {title} {Pairwise
  quantum correlations of a three-qubit xy chain with phase decoherence},\
  }\href {https://doi.org/10.1007/s11128-012-0460-1} {\bibfield  {journal}
  {\bibinfo  {journal} {Quantum Inf Process}\ }\textbf {\bibinfo {volume}
  {12}},\ \bibinfo {pages} {1141} (\bibinfo {year} {2013})}\BibitemShut
  {NoStop}%
\bibitem [{\citenamefont {Xie}\ and\ \citenamefont
  {Guo}(2013)}]{XieMeiQiu2013}%
  \BibitemOpen
  \bibfield  {author} {\bibinfo {author} {\bibfnamefont {M.-Q.}\ \bibnamefont
  {Xie}}\ and\ \bibinfo {author} {\bibfnamefont {B.}~\bibnamefont {Guo}},\
  }\bibfield  {title} {\bibinfo {title} {Thermal quantum discord in heisenberg
  {XXZ} model under different magnetic field conditions},\ }\href
  {https://doi.org/10.7498/aps.62.110303} {\bibfield  {journal} {\bibinfo
  {journal} {Acta Phys. Sin.}\ }\textbf {\bibinfo {volume} {62}},\ \bibinfo
  {pages} {110303} (\bibinfo {year} {2013})}\BibitemShut {NoStop}%
\bibitem [{\citenamefont {Fan}\ and\ \citenamefont
  {Zhang}(2013)}]{FanKaiMing2013}%
  \BibitemOpen
  \bibfield  {author} {\bibinfo {author} {\bibfnamefont {K.-M.}\ \bibnamefont
  {Fan}}\ and\ \bibinfo {author} {\bibfnamefont {G.-F.}\ \bibnamefont
  {Zhang}},\ }\bibfield  {title} {\bibinfo {title} {The dynamics of quantum
  correlation between two atoms in a damping {J}aynes--{C}ummings model},\
  }\href {https://doi.org/10.7498/aps.62.130301} {\bibfield  {journal}
  {\bibinfo  {journal} {Acta Phys. Sin.}\ }\textbf {\bibinfo {volume} {62}},\
  \bibinfo {pages} {130301} (\bibinfo {year} {2013})}\BibitemShut {NoStop}%
\bibitem [{\citenamefont {Li}\ and\ \citenamefont {Lu}(2014)}]{LiRuiQi2014}%
  \BibitemOpen
  \bibfield  {author} {\bibinfo {author} {\bibfnamefont {R.-Q.}\ \bibnamefont
  {Li}}\ and\ \bibinfo {author} {\bibfnamefont {D.-M.}\ \bibnamefont {Lu}},\
  }\bibfield  {title} {\bibinfo {title} {Quantum discord in the system of atoms
  interacting with coupled cavities},\ }\href
  {https://doi.org/10.7498/aps.63.030301} {\bibfield  {journal} {\bibinfo
  {journal} {Acta Phys. Sin.}\ }\textbf {\bibinfo {volume} {63}},\ \bibinfo
  {pages} {030301} (\bibinfo {year} {2014})}\BibitemShut {NoStop}%
\bibitem [{\citenamefont {Aldoshin}\ \emph {et~al.}(2014)\citenamefont
  {Aldoshin}, \citenamefont {Fel'dman},\ and\ \citenamefont
  {Yurishchev}}]{Aldoshin2014}%
  \BibitemOpen
  \bibfield  {author} {\bibinfo {author} {\bibfnamefont {S.~M.}\ \bibnamefont
  {Aldoshin}}, \bibinfo {author} {\bibfnamefont {E.~B.}\ \bibnamefont
  {Fel'dman}},\ and\ \bibinfo {author} {\bibfnamefont {M.~A.}\ \bibnamefont
  {Yurishchev}},\ }\bibfield  {title} {\bibinfo {title} {Quantum entanglement
  and quantum discord in magnetoactive materials (review article)},\ }\href
  {https://doi.org/10.1063/1.4862469} {\bibfield  {journal} {\bibinfo
  {journal} {Low Temperature Physics}\ }\textbf {\bibinfo {volume} {40}},\
  \bibinfo {pages} {3} (\bibinfo {year} {2014})}\BibitemShut {NoStop}%
\bibitem [{\citenamefont {Mohamed}(2018)}]{Mohamed2018}%
  \BibitemOpen
  \bibfield  {author} {\bibinfo {author} {\bibfnamefont {A.-B.~A.}\
  \bibnamefont {Mohamed}},\ }\bibfield  {title} {\bibinfo {title} {Bipartite
  non-classical correlations for a lossy two connected qubit–cavity systems:
  trace distance discord and bell’s non-locality},\ }\href
  {https://doi.org/10.1007/s11128-018-1865-2} {\bibfield  {journal} {\bibinfo
  {journal} {Quantum Inf Process}\ }\textbf {\bibinfo {volume} {17}} (\bibinfo
  {year} {2018})}\BibitemShut {NoStop}%
\bibitem [{\citenamefont {A.B.A.}\ \emph {et~al.}(2019)\citenamefont {A.B.A.},
  \citenamefont {H.},\ and\ \citenamefont {C.H.R.}}]{Mohamed2019}%
  \BibitemOpen
  \bibfield  {author} {\bibinfo {author} {\bibfnamefont {M.}~\bibnamefont
  {A.B.A.}}, \bibinfo {author} {\bibfnamefont {E.}~\bibnamefont {H.}},\ and\
  \bibinfo {author} {\bibfnamefont {O.}~\bibnamefont {C.H.R.}},\ }\bibfield
  {title} {\bibinfo {title} {Non-locality correlation in two driven qubits
  inside an open coherent cavity: Trace norm distance and maximum bell
  function},\ }\href {https://doi.org/10.1038/s41598-019-55548-2} {\bibfield
  {journal} {\bibinfo  {journal} {Sci Rep}\ }\textbf {\bibinfo {volume} {9}},\
  \bibinfo {pages} {19632} (\bibinfo {year} {2019})}\BibitemShut {NoStop}%
\bibitem [{\citenamefont {Jia}\ \emph {et~al.}(2020)\citenamefont {Jia},
  \citenamefont {Zhai}, \citenamefont {Yu}, \citenamefont {Wu},\ and\
  \citenamefont {Guo}}]{JiaZhihAhn2020}%
  \BibitemOpen
  \bibfield  {author} {\bibinfo {author} {\bibfnamefont {Z.-A.}\ \bibnamefont
  {Jia}}, \bibinfo {author} {\bibfnamefont {R.}~\bibnamefont {Zhai}}, \bibinfo
  {author} {\bibfnamefont {S.}~\bibnamefont {Yu}}, \bibinfo {author}
  {\bibfnamefont {Y.-C.}\ \bibnamefont {Wu}},\ and\ \bibinfo {author}
  {\bibfnamefont {G.-C.}\ \bibnamefont {Guo}},\ }\bibfield  {title} {\bibinfo
  {title} {Hierarchy of genuine multipartite quantum correlations},\ }\href
  {https://doi.org/10.1007/s11128-020-02922-z} {\bibfield  {journal} {\bibinfo
  {journal} {Quantum Inf Process}\ }\textbf {\bibinfo {volume} {19}},\ \bibinfo
  {pages} {419} (\bibinfo {year} {2020})}\BibitemShut {NoStop}%
\bibitem [{\citenamefont {Radhakrishnan}\ \emph {et~al.}(2020)\citenamefont
  {Radhakrishnan}, \citenamefont {Lauri\`ere},\ and\ \citenamefont
  {Byrnes}}]{Radhakrishnan2020}%
  \BibitemOpen
  \bibfield  {author} {\bibinfo {author} {\bibfnamefont {C.}~\bibnamefont
  {Radhakrishnan}}, \bibinfo {author} {\bibfnamefont {M.}~\bibnamefont
  {Lauri\`ere}},\ and\ \bibinfo {author} {\bibfnamefont {T.}~\bibnamefont
  {Byrnes}},\ }\bibfield  {title} {\bibinfo {title} {Multipartite
  generalization of quantum discord},\ }\href
  {https://doi.org/10.1103/PhysRevLett.124.110401} {\bibfield  {journal}
  {\bibinfo  {journal} {Phys. Rev. Lett.}\ }\textbf {\bibinfo {volume} {124}},\
  \bibinfo {pages} {110401} (\bibinfo {year} {2020})}\BibitemShut {NoStop}%
\bibitem [{\citenamefont {Brown}\ \emph {et~al.}(2021)\citenamefont {Brown},
  \citenamefont {Fawzi},\ and\ \citenamefont {Fawzi}}]{Brown2021}%
  \BibitemOpen
  \bibfield  {author} {\bibinfo {author} {\bibfnamefont {P.}~\bibnamefont
  {Brown}}, \bibinfo {author} {\bibfnamefont {H.}~\bibnamefont {Fawzi}},\ and\
  \bibinfo {author} {\bibfnamefont {O.}~\bibnamefont {Fawzi}},\ }\bibfield
  {title} {\bibinfo {title} {Computing conditional entropies for quantum
  correlations},\ }\href {https://doi.org/10.1038/s41467-020-20018-1}
  {\bibfield  {journal} {\bibinfo  {journal} {Nat Commun}\ }\textbf {\bibinfo
  {volume} {12}},\ \bibinfo {pages} {575} (\bibinfo {year} {2021})}\BibitemShut
  {NoStop}%
\bibitem [{\citenamefont {Jaynes}\ and\ \citenamefont
  {Cummings}(1963)}]{Jaynes1963}%
  \BibitemOpen
  \bibfield  {author} {\bibinfo {author} {\bibfnamefont {E.}~\bibnamefont
  {Jaynes}}\ and\ \bibinfo {author} {\bibfnamefont {F.}~\bibnamefont
  {Cummings}},\ }\bibfield  {title} {\bibinfo {title} {Comparison of quantum
  and semiclassical radiation theories with application to the beam maser},\
  }\href {https://doi.org/10.1109/PROC.1963.1664} {\bibfield  {journal}
  {\bibinfo  {journal} {Proceedings of the IEEE}\ }\textbf {\bibinfo {volume}
  {51}},\ \bibinfo {pages} {89} (\bibinfo {year} {1963})}\BibitemShut {NoStop}%
\bibitem [{\citenamefont {Tavis}\ and\ \citenamefont
  {Cummings}(1968)}]{Tavis1968}%
  \BibitemOpen
  \bibfield  {author} {\bibinfo {author} {\bibfnamefont {M.}~\bibnamefont
  {Tavis}}\ and\ \bibinfo {author} {\bibfnamefont {F.~W.}\ \bibnamefont
  {Cummings}},\ }\bibfield  {title} {\bibinfo {title} {Exact solution for an
  {$N$}-molecule---radiation-field {H}amiltonian},\ }\href
  {https://doi.org/10.1103/PhysRev.170.379} {\bibfield  {journal} {\bibinfo
  {journal} {Phys. Rev.}\ }\textbf {\bibinfo {volume} {170}},\ \bibinfo {pages}
  {379} (\bibinfo {year} {1968})}\BibitemShut {NoStop}%
\bibitem [{\citenamefont {Angelakis}\ \emph {et~al.}(2007)\citenamefont
  {Angelakis}, \citenamefont {Santos},\ and\ \citenamefont
  {Bose}}]{Angelakis2007}%
  \BibitemOpen
  \bibfield  {author} {\bibinfo {author} {\bibfnamefont {D.~G.}\ \bibnamefont
  {Angelakis}}, \bibinfo {author} {\bibfnamefont {M.~F.}\ \bibnamefont
  {Santos}},\ and\ \bibinfo {author} {\bibfnamefont {S.}~\bibnamefont {Bose}},\
  }\bibfield  {title} {\bibinfo {title} {Photon-blockade-induced mott
  transitions and {$XY$} spin models in coupled cavity arrays},\ }\href
  {https://doi.org/10.1103/PhysRevA.76.031805} {\bibfield  {journal} {\bibinfo
  {journal} {Phys. Rev. A}\ }\textbf {\bibinfo {volume} {76}},\ \bibinfo
  {pages} {031805} (\bibinfo {year} {2007})}\BibitemShut {NoStop}%
\bibitem [{\citenamefont {Wei}\ \emph {et~al.}(2021)\citenamefont {Wei},
  \citenamefont {Zhang}, \citenamefont {Greschner}, \citenamefont {Scott},\
  and\ \citenamefont {Zhang}}]{WeiHuanhuan2021}%
  \BibitemOpen
  \bibfield  {author} {\bibinfo {author} {\bibfnamefont {H.}~\bibnamefont
  {Wei}}, \bibinfo {author} {\bibfnamefont {J.}~\bibnamefont {Zhang}}, \bibinfo
  {author} {\bibfnamefont {S.}~\bibnamefont {Greschner}}, \bibinfo {author}
  {\bibfnamefont {T.~C.}\ \bibnamefont {Scott}},\ and\ \bibinfo {author}
  {\bibfnamefont {W.}~\bibnamefont {Zhang}},\ }\bibfield  {title} {\bibinfo
  {title} {Quantum monte carlo study of superradiant supersolid of light in the
  extended {J}aynes--{C}ummings--{H}ubbard model},\ }\href
  {https://doi.org/10.1103/PhysRevB.103.184501} {\bibfield  {journal} {\bibinfo
   {journal} {Phys. Rev. B}\ }\textbf {\bibinfo {volume} {103}},\ \bibinfo
  {pages} {184501} (\bibinfo {year} {2021})}\BibitemShut {NoStop}%
\bibitem [{\citenamefont {Prasad}\ and\ \citenamefont
  {Martin}(2018)}]{Prasad2018}%
  \BibitemOpen
  \bibfield  {author} {\bibinfo {author} {\bibfnamefont {S.}~\bibnamefont
  {Prasad}}\ and\ \bibinfo {author} {\bibfnamefont {A.}~\bibnamefont
  {Martin}},\ }\bibfield  {title} {\bibinfo {title} {Effective three-body
  interactions in {J}aynes--{C}ummings--{H}ubbard systems},\ }\href
  {https://doi.org/10.1038/s41598-018-33907-9} {\bibfield  {journal} {\bibinfo
  {journal} {Sci Rep}\ }\textbf {\bibinfo {volume} {8}},\ \bibinfo {pages}
  {16253} (\bibinfo {year} {2018})}\BibitemShut {NoStop}%
\bibitem [{\citenamefont {Guo}\ \emph {et~al.}(2019)\citenamefont {Guo},
  \citenamefont {Greschner}, \citenamefont {Zhu},\ and\ \citenamefont
  {Zhang}}]{GuoLijuan2019}%
  \BibitemOpen
  \bibfield  {author} {\bibinfo {author} {\bibfnamefont {L.}~\bibnamefont
  {Guo}}, \bibinfo {author} {\bibfnamefont {S.}~\bibnamefont {Greschner}},
  \bibinfo {author} {\bibfnamefont {S.}~\bibnamefont {Zhu}},\ and\ \bibinfo
  {author} {\bibfnamefont {W.}~\bibnamefont {Zhang}},\ }\bibfield  {title}
  {\bibinfo {title} {Supersolid and pair correlations of the extended
  {J}aynes--{C}ummings--{H}ubbard model on triangular lattices},\ }\href
  {https://doi.org/10.1103/PhysRevA.100.033614} {\bibfield  {journal} {\bibinfo
   {journal} {Phys. Rev. A}\ }\textbf {\bibinfo {volume} {100}},\ \bibinfo
  {pages} {033614} (\bibinfo {year} {2019})}\BibitemShut {NoStop}%
\bibitem [{\citenamefont {Smith}\ \emph {et~al.}(2021)\citenamefont {Smith},
  \citenamefont {Bhattacharya},\ and\ \citenamefont {Masiello}}]{Smith2021}%
  \BibitemOpen
  \bibfield  {author} {\bibinfo {author} {\bibfnamefont {K.~C.}\ \bibnamefont
  {Smith}}, \bibinfo {author} {\bibfnamefont {A.}~\bibnamefont
  {Bhattacharya}},\ and\ \bibinfo {author} {\bibfnamefont {D.~J.}\ \bibnamefont
  {Masiello}},\ }\bibfield  {title} {\bibinfo {title} {Exact $k$-body
  representation of the {J}aynes--{C}ummings interaction in the dressed basis:
  Insight into many-body phenomena with light},\ }\href
  {https://doi.org/10.1103/PhysRevA.104.013707} {\bibfield  {journal} {\bibinfo
   {journal} {Phys. Rev. A}\ }\textbf {\bibinfo {volume} {104}},\ \bibinfo
  {pages} {013707} (\bibinfo {year} {2021})}\BibitemShut {NoStop}%
\bibitem [{\citenamefont {Düll}\ \emph {et~al.}(2021)\citenamefont {Düll},
  \citenamefont {Kulagin}, \citenamefont {Lee}, \citenamefont {Ozhigov},
  \citenamefont {Miao},\ and\ \citenamefont {Zheng}}]{Dull2021}%
  \BibitemOpen
  \bibfield  {author} {\bibinfo {author} {\bibfnamefont {R.}~\bibnamefont
  {Düll}}, \bibinfo {author} {\bibfnamefont {A.}~\bibnamefont {Kulagin}},
  \bibinfo {author} {\bibfnamefont {W.}~\bibnamefont {Lee}}, \bibinfo {author}
  {\bibfnamefont {Y.}~\bibnamefont {Ozhigov}}, \bibinfo {author} {\bibfnamefont
  {H.}~\bibnamefont {Miao}},\ and\ \bibinfo {author} {\bibfnamefont
  {K.}~\bibnamefont {Zheng}},\ }\bibfield  {title} {\bibinfo {title} {Quality
  of control in the {T}avis--{C}ummings--{H}ubbard model},\ }\href
  {https://doi.org/10.1007/s10598-021-09517-y} {\bibfield  {journal} {\bibinfo
  {journal} {Computational Mathematics and Modeling}\ }\textbf {\bibinfo
  {volume} {32}},\ \bibinfo {pages} {75} (\bibinfo {year} {2021})}\BibitemShut
  {NoStop}%
\bibitem [{\citenamefont {Afanasyev}\ \emph {et~al.}(2021)\citenamefont
  {Afanasyev}, \citenamefont {Zheng}, \citenamefont {Kulagin}, \citenamefont
  {Miao}, \citenamefont {Ozhigov}, \citenamefont {Lee},\ and\ \citenamefont
  {Victorova}}]{Afanasyev2021}%
  \BibitemOpen
  \bibfield  {author} {\bibinfo {author} {\bibfnamefont {V.}~\bibnamefont
  {Afanasyev}}, \bibinfo {author} {\bibfnamefont {K.}~\bibnamefont {Zheng}},
  \bibinfo {author} {\bibfnamefont {A.}~\bibnamefont {Kulagin}}, \bibinfo
  {author} {\bibfnamefont {H.}~\bibnamefont {Miao}}, \bibinfo {author}
  {\bibfnamefont {Y.}~\bibnamefont {Ozhigov}}, \bibinfo {author} {\bibfnamefont
  {W.}~\bibnamefont {Lee}},\ and\ \bibinfo {author} {\bibfnamefont
  {N.}~\bibnamefont {Victorova}},\ }\bibfield  {title} {\bibinfo {title} {About
  chemical modifications of finite dimensional {QED} models},\ }\href
  {https://doi.org/10.33581/1561-4085-2021-24-3-230-241} {\bibfield  {journal}
  {\bibinfo  {journal} {Nonlinear Phenomena in Complex Systems}\ }\textbf
  {\bibinfo {volume} {24}},\ \bibinfo {pages} {230} (\bibinfo {year}
  {2021})}\BibitemShut {NoStop}%
\bibitem [{\citenamefont {hui Miao}\ and\ \citenamefont
  {Ozhigov}(2023)}]{Miao2023}%
  \BibitemOpen
  \bibfield  {author} {\bibinfo {author} {\bibfnamefont {H.}~\bibnamefont {hui
  Miao}}\ and\ \bibinfo {author} {\bibfnamefont {Y.~I.}\ \bibnamefont
  {Ozhigov}},\ }\bibfield  {title} {\bibinfo {title} {Using a modified version
  of the {T}avis--{C}ummings--{H}ubbard model to simulate the formation of
  neutral hydrogen molecule},\ }\href
  {https://doi.org/https://doi.org/10.1016/j.physa.2023.128851} {\bibfield
  {journal} {\bibinfo  {journal} {Physica A: Statistical Mechanics and its
  Applications}\ }\textbf {\bibinfo {volume} {622}},\ \bibinfo {pages} {128851}
  (\bibinfo {year} {2023})}\BibitemShut {NoStop}%
\bibitem [{\citenamefont {Miao}\ and\ \citenamefont
  {Ozhigov}(2023)}]{Ozhigov2023}%
  \BibitemOpen
  \bibfield  {author} {\bibinfo {author} {\bibfnamefont {H.-h.}\ \bibnamefont
  {Miao}}\ and\ \bibinfo {author} {\bibfnamefont {Y.~I.}\ \bibnamefont
  {Ozhigov}},\ }\bibfield  {title} {\bibinfo {title} {Comparing the effects of
  nuclear and electron spins on the formation of neutral hydrogen molecule},\
  }\href {https://doi.org/10.1134/S1995080223080401} {\bibfield  {journal}
  {\bibinfo  {journal} {Lobachevskii Journal of Mathematics}\ }\textbf
  {\bibinfo {volume} {44}},\ \bibinfo {pages} {3111} (\bibinfo {year}
  {2023})}\BibitemShut {NoStop}%
\bibitem [{\citenamefont {Miao}\ and\ \citenamefont
  {Ozhigov}(2024)}]{Miaohuihui2024}%
  \BibitemOpen
  \bibfield  {author} {\bibinfo {author} {\bibfnamefont {H.-h.}\ \bibnamefont
  {Miao}}\ and\ \bibinfo {author} {\bibfnamefont {Y.~I.}\ \bibnamefont
  {Ozhigov}},\ }\bibfield  {title} {\bibinfo {title} {Distributed computing
  quantum unitary evolution},\ }\href
  {https://doi.org/10.1134/S1995080224603904} {\bibfield  {journal} {\bibinfo
  {journal} {Lobachevskii Journal of Mathematics}\ }\textbf {\bibinfo {volume}
  {45}},\ \bibinfo {pages} {3130} (\bibinfo {year} {2024})}\BibitemShut
  {NoStop}%
\bibitem [{\citenamefont {Li}\ \emph {et~al.}(2024)\citenamefont {Li},
  \citenamefont {Miao},\ and\ \citenamefont {Ozhigov}}]{Li2024}%
  \BibitemOpen
  \bibfield  {author} {\bibinfo {author} {\bibfnamefont {W.}~\bibnamefont
  {Li}}, \bibinfo {author} {\bibfnamefont {H.-h.}\ \bibnamefont {Miao}},\ and\
  \bibinfo {author} {\bibfnamefont {Y.~I.}\ \bibnamefont {Ozhigov}},\
  }\bibfield  {title} {\bibinfo {title} {Supercomputer model of
  finite-dimensional quantum electrodynamics applications},\ }\href
  {https://doi.org/10.1134/S1995080224603849} {\bibfield  {journal} {\bibinfo
  {journal} {Lobachevskii Journal of Mathematics}\ }\textbf {\bibinfo {volume}
  {45}},\ \bibinfo {pages} {3106} (\bibinfo {year} {2024})}\BibitemShut
  {NoStop}%
\bibitem [{\citenamefont {Obada}\ \emph {et~al.}(2007)\citenamefont {Obada},
  \citenamefont {Hessian},\ and\ \citenamefont {Mohamed}}]{Obada2007}%
  \BibitemOpen
  \bibfield  {author} {\bibinfo {author} {\bibfnamefont {A.-S.~F.}\
  \bibnamefont {Obada}}, \bibinfo {author} {\bibfnamefont {H.~A.}\ \bibnamefont
  {Hessian}},\ and\ \bibinfo {author} {\bibfnamefont {A.-B.~A.}\ \bibnamefont
  {Mohamed}},\ }\bibfield  {title} {\bibinfo {title} {Entropies and
  entanglement for decoherence without energy relaxation in a two-level atom},\
  }\href {https://doi.org/10.1088/0953-4075/40/12/002} {\bibfield  {journal}
  {\bibinfo  {journal} {Journal of Physics B: Atomic, Molecular and Optical
  Physics}\ }\textbf {\bibinfo {volume} {40}},\ \bibinfo {pages} {2241}
  (\bibinfo {year} {2007})}\BibitemShut {NoStop}%
\bibitem [{\citenamefont {Obada}\ \emph {et~al.}(2008)\citenamefont {Obada},
  \citenamefont {Hessian},\ and\ \citenamefont {Mohamed}}]{Obada2008}%
  \BibitemOpen
  \bibfield  {author} {\bibinfo {author} {\bibfnamefont {A.-S.}\ \bibnamefont
  {Obada}}, \bibinfo {author} {\bibfnamefont {H.}~\bibnamefont {Hessian}},\
  and\ \bibinfo {author} {\bibfnamefont {A.-B.}\ \bibnamefont {Mohamed}},\
  }\bibfield  {title} {\bibinfo {title} {Effect of phase-damped cavity on
  dynamics of tangles of a nondegenerate two-photon jc model},\ }\href
  {https://doi.org/https://doi.org/10.1016/j.optcom.2008.06.076} {\bibfield
  {journal} {\bibinfo  {journal} {Optics Communications}\ }\textbf {\bibinfo
  {volume} {281}},\ \bibinfo {pages} {5189} (\bibinfo {year}
  {2008})}\BibitemShut {NoStop}%
\bibitem [{\citenamefont {Bennett}\ \emph {et~al.}(1996)\citenamefont
  {Bennett}, \citenamefont {Bernstein}, \citenamefont {Popescu},\ and\
  \citenamefont {Schumacher}}]{Bennett1996}%
  \BibitemOpen
  \bibfield  {author} {\bibinfo {author} {\bibfnamefont {C.~H.}\ \bibnamefont
  {Bennett}}, \bibinfo {author} {\bibfnamefont {H.~J.}\ \bibnamefont
  {Bernstein}}, \bibinfo {author} {\bibfnamefont {S.}~\bibnamefont {Popescu}},\
  and\ \bibinfo {author} {\bibfnamefont {B.}~\bibnamefont {Schumacher}},\
  }\bibfield  {title} {\bibinfo {title} {Concentrating partial entanglement by
  local operations},\ }\href {https://doi.org/10.1103/PhysRevA.53.2046}
  {\bibfield  {journal} {\bibinfo  {journal} {Phys. Rev. A}\ }\textbf {\bibinfo
  {volume} {53}},\ \bibinfo {pages} {2046} (\bibinfo {year}
  {1996})}\BibitemShut {NoStop}%
\bibitem [{\citenamefont {Hill}\ and\ \citenamefont
  {Wootters}(1997)}]{Hill1997}%
  \BibitemOpen
  \bibfield  {author} {\bibinfo {author} {\bibfnamefont {S.~A.}\ \bibnamefont
  {Hill}}\ and\ \bibinfo {author} {\bibfnamefont {W.~K.}\ \bibnamefont
  {Wootters}},\ }\bibfield  {title} {\bibinfo {title} {Entanglement of a pair
  of quantum bits},\ }\href {https://doi.org/10.1103/PhysRevLett.78.5022}
  {\bibfield  {journal} {\bibinfo  {journal} {Phys. Rev. Lett.}\ }\textbf
  {\bibinfo {volume} {78}},\ \bibinfo {pages} {5022} (\bibinfo {year}
  {1997})}\BibitemShut {NoStop}%
\bibitem [{\citenamefont {Wootters}(1998)}]{Wootters1998}%
  \BibitemOpen
  \bibfield  {author} {\bibinfo {author} {\bibfnamefont {W.~K.}\ \bibnamefont
  {Wootters}},\ }\bibfield  {title} {\bibinfo {title} {Entanglement of
  formation of an arbitrary state of two qubits},\ }\href
  {https://doi.org/10.1103/PhysRevLett.80.2245} {\bibfield  {journal} {\bibinfo
   {journal} {Phys. Rev. Lett.}\ }\textbf {\bibinfo {volume} {80}},\ \bibinfo
  {pages} {2245} (\bibinfo {year} {1998})}\BibitemShut {NoStop}%
\bibitem [{\citenamefont {Audenaert}\ \emph {et~al.}(2001)\citenamefont
  {Audenaert}, \citenamefont {Verstraete},\ and\ \citenamefont
  {De~Moor}}]{Audenaert2001}%
  \BibitemOpen
  \bibfield  {author} {\bibinfo {author} {\bibfnamefont {K.}~\bibnamefont
  {Audenaert}}, \bibinfo {author} {\bibfnamefont {F.}~\bibnamefont
  {Verstraete}},\ and\ \bibinfo {author} {\bibfnamefont {B.}~\bibnamefont
  {De~Moor}},\ }\bibfield  {title} {\bibinfo {title} {Variational
  characterizations of separability and entanglement of formation},\ }\href
  {https://doi.org/10.1103/PhysRevA.64.052304} {\bibfield  {journal} {\bibinfo
  {journal} {Phys. Rev. A}\ }\textbf {\bibinfo {volume} {64}},\ \bibinfo
  {pages} {052304} (\bibinfo {year} {2001})}\BibitemShut {NoStop}%
\bibitem [{\citenamefont {Wang}\ and\ \citenamefont
  {Zanardi}(2002)}]{Wang2002}%
  \BibitemOpen
  \bibfield  {author} {\bibinfo {author} {\bibfnamefont {X.}~\bibnamefont
  {Wang}}\ and\ \bibinfo {author} {\bibfnamefont {P.}~\bibnamefont {Zanardi}},\
  }\bibfield  {title} {\bibinfo {title} {Quantum entanglement and bell
  inequalities in heisenberg spin chains},\ }\href
  {https://doi.org/https://doi.org/10.1016/S0375-9601(02)00885-X} {\bibfield
  {journal} {\bibinfo  {journal} {Physics Letters A}\ }\textbf {\bibinfo
  {volume} {301}},\ \bibinfo {pages} {1} (\bibinfo {year} {2002})}\BibitemShut
  {NoStop}%
\bibitem [{\citenamefont {Laflorencie}(2016)}]{Laflorencie2016}%
  \BibitemOpen
  \bibfield  {author} {\bibinfo {author} {\bibfnamefont {N.}~\bibnamefont
  {Laflorencie}},\ }\bibfield  {title} {\bibinfo {title} {Quantum entanglement
  in condensed matter systems},\ }\href
  {https://doi.org/https://doi.org/10.1016/j.physrep.2016.06.008} {\bibfield
  {journal} {\bibinfo  {journal} {Physics Reports}\ }\textbf {\bibinfo {volume}
  {646}},\ \bibinfo {pages} {1} (\bibinfo {year} {2016})},\ \bibinfo {note}
  {quantum entanglement in condensed matter systems}\BibitemShut {NoStop}%
\bibitem [{\citenamefont {L.S.}(2021)}]{Lima2021}%
  \BibitemOpen
  \bibfield  {author} {\bibinfo {author} {\bibfnamefont {L.}~\bibnamefont
  {L.S.}},\ }\bibfield  {title} {\bibinfo {title} {Quantum correlation and
  entanglement in the heisenberg model with biquadratic interaction on square
  lattice},\ }\href {https://doi.org/10.1140/epjd/s10053-021-00044-4}
  {\bibfield  {journal} {\bibinfo  {journal} {Eur. Phys. J. D}\ }\textbf
  {\bibinfo {volume} {75}} (\bibinfo {year} {2021})}\BibitemShut {NoStop}%
\bibitem [{\citenamefont {Wu}\ and\ \citenamefont {Yang}(2007)}]{Wu2007}%
  \BibitemOpen
  \bibfield  {author} {\bibinfo {author} {\bibfnamefont {Y.}~\bibnamefont
  {Wu}}\ and\ \bibinfo {author} {\bibfnamefont {X.}~\bibnamefont {Yang}},\
  }\bibfield  {title} {\bibinfo {title} {Strong-coupling theory of periodically
  driven two-level systems},\ }\href
  {https://doi.org/10.1103/PhysRevLett.98.013601} {\bibfield  {journal}
  {\bibinfo  {journal} {Phys. Rev. Lett.}\ }\textbf {\bibinfo {volume} {98}},\
  \bibinfo {pages} {013601} (\bibinfo {year} {2007})}\BibitemShut {NoStop}%
\bibitem [{\citenamefont {Breuer}\ \emph {et~al.}(2002)\citenamefont {Breuer},
  \citenamefont {Petruccione} \emph {et~al.}}]{Breuer2002}%
  \BibitemOpen
  \bibfield  {author} {\bibinfo {author} {\bibfnamefont {H.-P.}\ \bibnamefont
  {Breuer}}, \bibinfo {author} {\bibfnamefont {F.}~\bibnamefont {Petruccione}},
  \emph {et~al.},\ }\href
  {http://refhub.elsevier.com/S0378-4371(22)00557-X/sb42} {\emph {\bibinfo
  {title} {The theory of open quantum systems}}}\ (\bibinfo  {publisher}
  {Oxford University Press},\ \bibinfo {year} {2002})\BibitemShut {NoStop}%
\bibitem [{\citenamefont {Alicki}(1979)}]{Alicki1979}%
  \BibitemOpen
  \bibfield  {author} {\bibinfo {author} {\bibfnamefont {R.}~\bibnamefont
  {Alicki}},\ }\bibfield  {title} {\bibinfo {title} {The quantum open system as
  a model of the heat engine},\ }\href
  {https://doi.org/10.1088/0305-4470/12/5/007} {\bibfield  {journal} {\bibinfo
  {journal} {Journal of Physics A: Mathematical and General}\ }\textbf
  {\bibinfo {volume} {12}},\ \bibinfo {pages} {L103} (\bibinfo {year}
  {1979})}\BibitemShut {NoStop}%
\bibitem [{\citenamefont {Kosloff}(2013)}]{Kosloff2013}%
  \BibitemOpen
  \bibfield  {author} {\bibinfo {author} {\bibfnamefont {R.}~\bibnamefont
  {Kosloff}},\ }\bibfield  {title} {\bibinfo {title} {Quantum thermodynamics: A
  dynamical viewpoint},\ }\href {https://doi.org/10.3390/e15062100} {\bibfield
  {journal} {\bibinfo  {journal} {Entropy}\ }\textbf {\bibinfo {volume} {15}},\
  \bibinfo {pages} {2100} (\bibinfo {year} {2013})}\BibitemShut {NoStop}%
\bibitem [{\citenamefont {Rabi}(1936)}]{Rabi1936}%
  \BibitemOpen
  \bibfield  {author} {\bibinfo {author} {\bibfnamefont {I.~I.}\ \bibnamefont
  {Rabi}},\ }\bibfield  {title} {\bibinfo {title} {On the process of space
  quantization},\ }\href {https://doi.org/10.1103/PhysRev.49.324} {\bibfield
  {journal} {\bibinfo  {journal} {Phys. Rev.}\ }\textbf {\bibinfo {volume}
  {49}},\ \bibinfo {pages} {324} (\bibinfo {year} {1936})}\BibitemShut
  {NoStop}%
\end{thebibliography}%

\end{document}